\documentclass[11pt]{article}
\usepackage{epsfig,graphicx,cite, mathrsfs}
\usepackage[percent]{overpic}
\input amssym.def
\input amssym.tex
\let\svthefootnote\thefootnote
\newcommand\blankfootnote[1]{%
  \let\thefootnote\relax\footnotetext{#1}%
  \let\thefootnote\svthefootnote%
}

\makeatletter
\@addtoreset{equation}{section}
\makeatother

\def\ben{\begin{equation}}
\def\een{\end{equation}}

\def\bea{\begin{eqnarray}}
\def\eea{\end{eqnarray}}

\def\mathbb{\Bbb}
\def\be{\begin{equation}}
\def\ee{\end{equation}}

\newcount\hour \newcount\minute
\hour=\time  \divide \hour by 60
\minute=\time

\def\mathbb{\Bbb}

\newcount\hour \newcount\minute
\hour=\time  \divide \hour by 60
\minute=\time
\loop \ifnum \minute > 59 \advance \minute by -60 \repeat
\def\nowtwelve{\ifnum \hour<13 \number\hour:
                      \ifnum \minute<10 0\fi
                      \number\minute
                      \ifnum \hour<12 \ A.M.\else \ P.M.\fi
         \else \advance \hour by -12 \number\hour:
                      \ifnum \minute<10 0\fi
                      \number\minute \ P.M.\fi}
\def\nowtwentyfour{\ifnum \hour<10 0\fi
                \number\hour:
                \ifnum \minute<10 0\fi
                \number\minute}
\def\now{\nowtwelve}
\newcommand{\hoch}[1]{$\, ^{#1}$}

\newcommand{\auth}{\Large\bf{ 
G.W. Gibbons\hoch{1,2} and C.M. Warnick\hoch{3}
}}

\begin{document}


\begin{center}

{\LARGE{\bf Aspherical Photon and Anti-Photon Surfaces }}

\vspace{12pt}

\auth

\large
\vspace{3pt}{\hoch{1}\it DAMTP, Centre for Mathematical Sciences,\\
 Cambridge University, Wilberforce Road, Cambridge CB3 OWA, UK}

\vspace{0pt}{\hoch{2}\it Laboratoire de Math\'ematiques et Physique Th\'eorique, CNRS-UMR 7350, Universit\'e de Tours, Parc de Grandmont, 37200 Tours, France}

\vspace{3pt}{\hoch{3}\it Department of Mathematics,
South Kensington Campus,\\
Imperial College London, \,
London, SW7 2AZ, UK }

\end{center}

\newcount\hour \newcount\minute
\hour=\time  \divide \hour by 60
\minute=\time
\loop \ifnum \minute > 59 \advance \minute by -60 \repeat
\def\nowtwelve{\ifnum \hour<13 \number\hour:
                      \ifnum \minute<10 0\fi
                      \number\minute
                      \ifnum \hour<12 \ A.M.\else \ P.M.\fi
         \else \advance \hour by -12 \number\hour:
                      \ifnum \minute<10 0\fi
                      \number\minute \ P.M.\fi}
\def\nowtwentyfour{\ifnum \hour<10 0\fi
                \number\hour:
                \ifnum \minute<10 0\fi
                \number\minute}
\def\now{\nowtwelve}

\medskip
\centerline{\today}

\begin{abstract}
In this note we identify photon surfaces and anti-photon surfaces in some physically interesting spacetimes, which are not spherically symmetric. All of our examples solve physically reasonable field equations, including for some cases the vacuum Einstein equations, albeit they are not asymptotically flat. Our examples include the vacuum C-metric, the Melvin solution of Einstein-Maxwell theory and generalisations including dilaton fields. The (anti-)photon surfaces are not round spheres, and the lapse function is not always constant. 
\end{abstract}

\blankfootnote{e-mail: \texttt{g.w.gibbons@damtp.cam.ac.uk}; \texttt{c.warnick@imperial.ac.uk}}




\section{Introduction}

It is well known that the Schwarzschild solution
contains  circular photon orbits  at $r=3M$, where  $M>0$ 
is the ADM mass. These circular  photon orbits are the projection 
onto the spatial manifold $t={\rm constant}$  of null geodesics
in the spacetime. Moreover if the projection of the tangent vector of  
any null geodesic is tangent to the sphere  at one time
it remains tangent to the sphere at all future times. Because
the Schwarzschild metric is static it is both  possible 
and convenient to 
reformulate these properties using Fermat's principle
in terms of the so-called  optical   geometry
of the spatial sections. Any static spacetime metric may cast in the form
\ben
ds^2 = g_{\mu \nu} dx^\mu dx^\nu=  -N^2 dt ^2 + g_{ij}dx^idx^j
\een
with $x^\mu = (t,x^i)$, $i=1,2,3$ and the lapse function $N$ and 
spatial metric $g_{ij}$  independent of $t$.
It is a straightforward exercise to show that the spatial projection
of null geodesics are geodesics of the optical distance $ds_{\rm opt}$ defined 
by
\ben
ds^2 _{\rm opt} = N^{-2} g_{ij}dx^idx^j  = f_{ij} dx^i dx^j
\,.\een 
For the Schwarzschild solution
\ben
 ds^2 _{\rm opt} = \frac{dr^2}{\left (1-\frac{2M}{r}\right)^2 } + 
\frac{r^2 }{1-\frac{2M}{r} } 
( d \theta ^2 + \sin ^2 \theta d \phi ^2 ) \,.
\een
The circumference  $C(r)$ of every great circle lying on 
the sphere $r={\rm constant}$ is given by
\ben
C(r)= \frac{2 \pi r}{\sqrt{1-\frac{2M}{r}}}.  
\een
The circumference $C(r)$ has a unique minimum at
$r=3M$. Thus every great circle lying on the sphere $r=3M$ 
is a geodesic  of the ambient three-dimensional optical manifold.
Expressed differently: $r=3M$ is a totally geodesic submanifold 
(in fact hypersurface) of the optical manifold.
 
Photon surfaces have attracted attention recently, in particular in the last two years there have been several results establishing the uniqueness of spacetimes admitting a photon surface under certain conditions \cite{Cederbaum:2014gva, Cederbaum:2015aha, Cederbaum:2015fra, Yazadjiev:2015mta, Yazadjiev:2015hda, Yazadjiev:2015jza, Rogatko:2016mho, Yoshino:2016kgi}. These works typically assume that the spacetime is complete, asymptotically flat and with the exception of \cite{Yoshino:2016kgi} assume  that the lapse, $N$, is constant on the surface.  In this paper we give some counter-examples to demonstrate that the conclusions of these theorems can be violated if one allows certain of the assumptions to be dropped. In particular, we shall show that there exist physically interesting metrics satisfying Einstein's equations (with or without matter) with non-spherically symmetric photon spheres such that the lapse is not constant on the photon sphere. Moreover these metrics are not of cohomogeneity one. The metrics contain relatively mild (conical) singularities, and are not asymptotically flat in the usual sense (although in the $\Lambda=0$ case they contain regions in which the curvature  approaches zero). These spacetimes we consider are all related to the C-metrics, first found by Levi-Civita \cite{Levi-Civita}, which are now 
understood to represent uniformly accelerated black holes.

Anti-photon surfaces are much less well known. They correspond in the static setting to totally geodesic submanifolds of the optical metric for which, however, the photon orbits lying in the surface are stable (as opposed to the unstable case characterising the photon surfaces). In the spherically symmetric case, in the absence of naked singularities, it seems that these cannot occur if the energy-momentum tensor satisfies reasonable energy conditions \cite{Cvetic:2016bxi}. However, in a class of cylindrically symmetric spacetimes of Melvin type \cite{Melvin:1965zza} we present anti-photon cylinders.

\section{Some  Aspherical photon spheres}

\subsection{ The Vacuum C-metric}

While the existence of a photon surface
surrounding a spherically symmetric black hole is 
not surprising, the fact that it persists
when  the black hole  undergoes  a uniform acceleration and ceases to be 
spherically symmetric is not at all obvious. This 
situation is described by the `C-metric' first found by Levi-Civita
\cite{Levi-Civita}. Its physical significance was 
first elucidated by Kinnersly and Walker  \cite{Kinnersley:1970zw,KW2}.
For a subsequent review see \cite{CornishUttley}. For a uniqueness
theorem see  \cite{Wells:1998gb}.

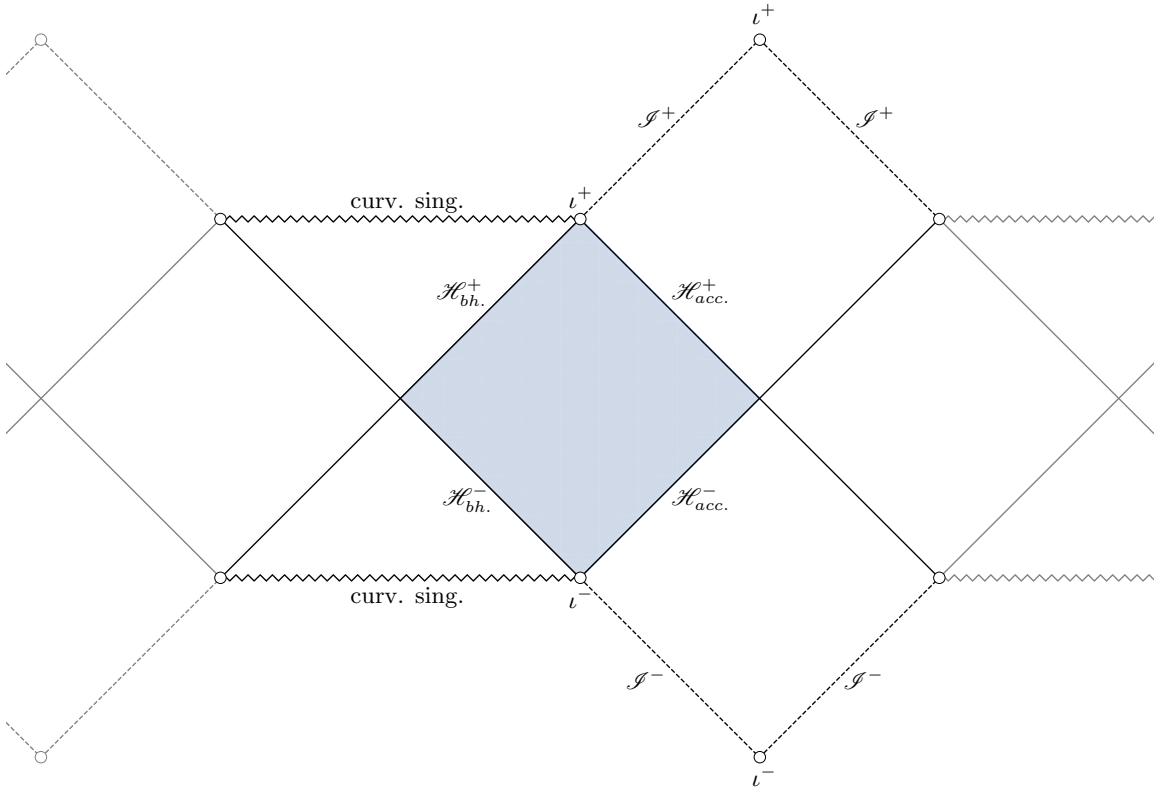
\begin{figure}[ht]
\vspace{.3cm}
\begin{overpic}[width=\textwidth]{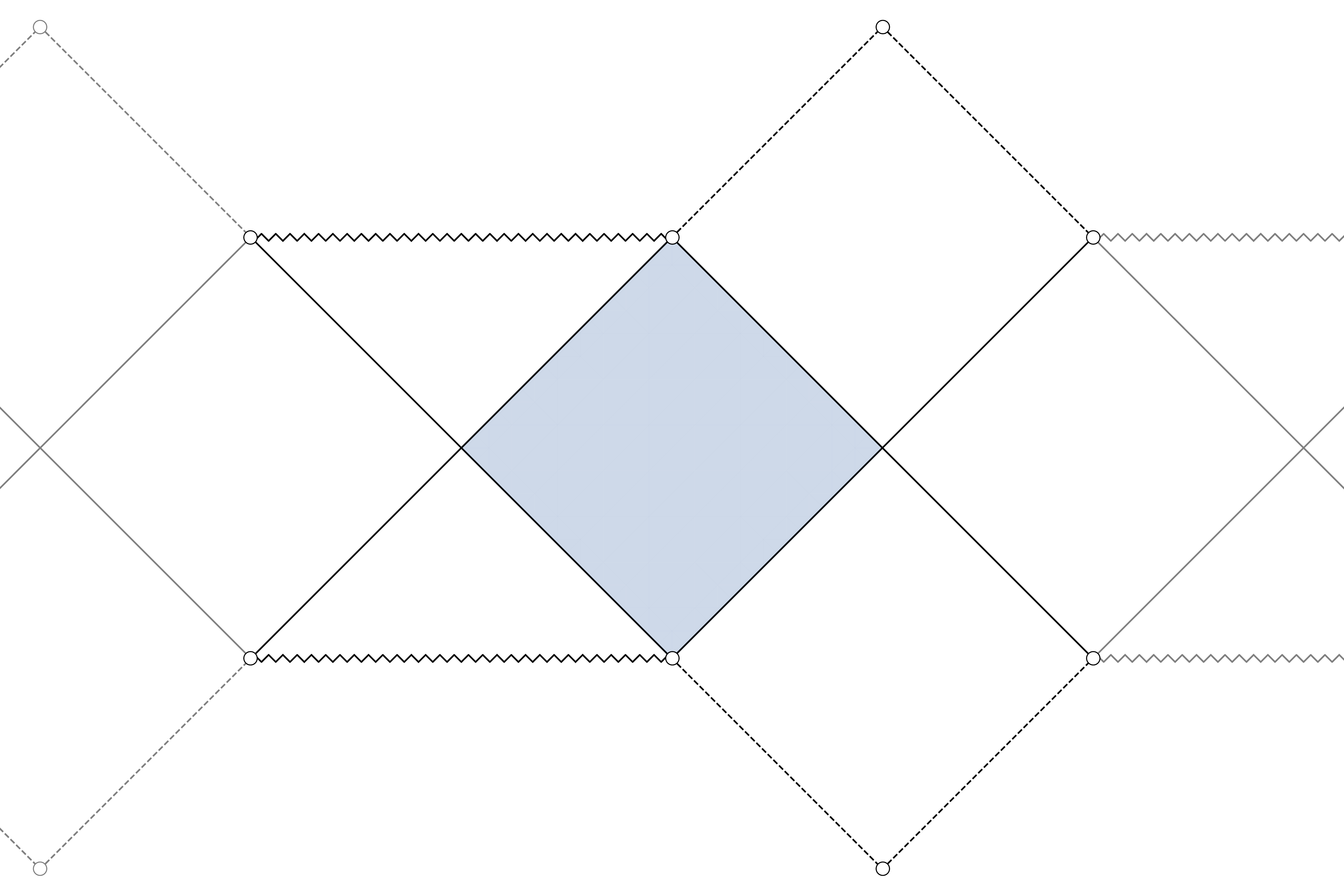}
\put (74, 57) {\footnotesize$\mathscr{I}^+$}
\put (55, 57) {\footnotesize$\mathscr{I}^+$}
\put (65, 66) {\footnotesize$\iota^+$}
\put (49, 50) {\footnotesize$\iota^+$}
\put (37.5, 42) {\footnotesize$\mathscr{H}_{bh.}^+$}
\put (58, 42) {\footnotesize$\mathscr{H}_{acc.}^+$}
\put (38, 24) {\footnotesize$\mathscr{H}_{bh.}^-$}
\put (58, 24) {\footnotesize$\mathscr{H}_{acc.}^-$}
\put (65, -.5) {\footnotesize$\iota^-$}
\put (49, 15) {\footnotesize$\iota^-$}
\put (73, 8) {\footnotesize$\mathscr{I}^-$}
\put (54, 8) {\footnotesize$\mathscr{I}^-$}
\put (30, 50) {\footnotesize curv. sing.}
\put (30, 15.5) {\footnotesize curv. sing.}
\end{overpic}
\caption{A section of the Penrose diagram of the maximally analytically extended uncharged C-metric without cosmological constant. The shaded region corresponds to a static patch. \label{penrose}}
\end{figure}
 
The metric is  given in  in Hong-Teo coordinates \cite{Hong:2003gx} by 
\ben ds ^2  = \frac{1}{a^2(x+y)^2}\left ( F(y) dt^2 - \frac{1}{F(y)}dy^2 +
  \frac{1}{F(x)} dx^2 +F(x) d\phi^2 \right) \label{cmetric}
\,, \een
where 
\begin{equation}
F(u) = (1-u^2)(1+2 mau).
\end{equation}
$F(y)$ is negative on the interval $(-1/2ma, -1)$
and the metric is static in this region, with Killing horizons at
$-1/2ma$, $-1$ corresponding to a black hole horizon and an acceleration
horizon respectively. The coordinate $x$ takes values in $(-1,1)$ and for $a \neq 0$, there will in general be conical
singularities on the axis $x=\pm 1$. Choosing the period of $\phi$,
one can eliminate the singularity on either $x=1$ or $x=-1$. We
can interpret the singularity as either representing a strut pushing the
black hole or else a string pulling it depending on which choice we make.

In Figure \ref{penrose} we show the Penrose diagram of the maximally extended C-metric. The shaded region in the figure corresponds to the region $-1/2ma<y< -1$, and the two Killing horizons are shown. Each point in the interior of the shaded region represents a topological sphere with coordinates $x, \phi$. This sphere is not round, but is axisymmetric and further has at least one conical singularity on the axis (see Figure \ref{cmetps} for an embedded example).  The spacetime has an asymptotic region which is accessible from the static region by causal curves falling through the acceleration horizon. This region is asymptotically flat in the sense that the curvature decays along causal curves.

The optical metric is given by
\ben ds_{\rm opt} = \frac{1}{F(y)^2}dy^2 + \frac{1}{\left |F(y)\right |} 
\left( \frac{dx^2}{F(x)} + F(x) d\phi^2 \right).
\een
 Since $\left |F(y)\right |$ vanishes at the black hole horizon
and the acceleration horizon, it must have at least one maximum on the
interval $(-1/2ma, -1)$. This corresponds to a photon surface, and 
furthermore it is unstable, in the sense that geodesics which start close to the surface do not remain so. This surface will
generically have a conical singularity corresponding to that of the
full C-metric. In Figure \ref{cmetps} we show an isometric embedding
of the C-metric photon surface into Euclidean space. We identify $\phi$
so that the acceleration is induced by a string in this example (the
other case does not allow an embedding into flat space).

\begin{figure}[ht]
\begin{center}
\includegraphics[height=5cm]{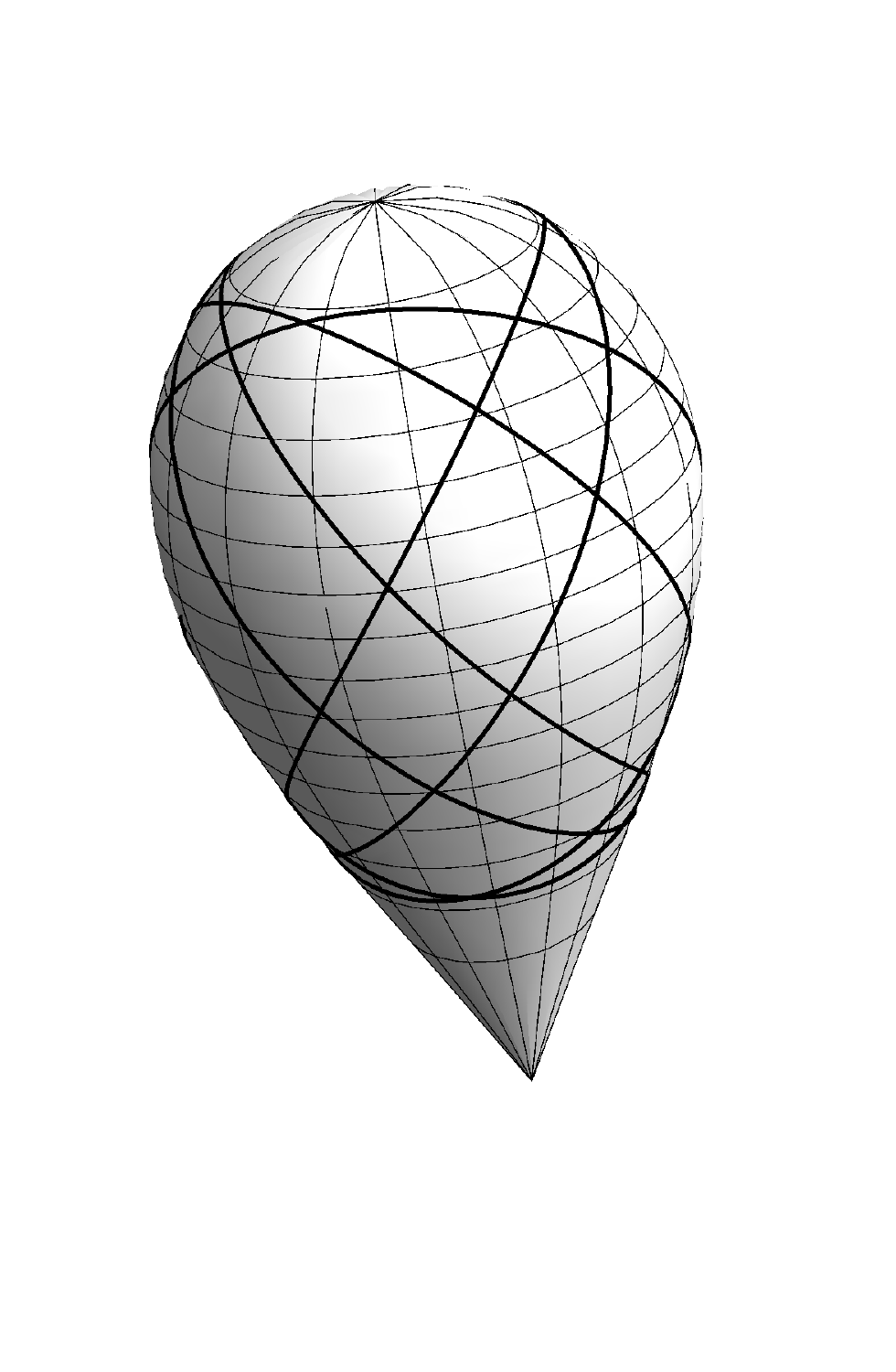}
\caption{The photon surface for the C-metric with $ma=0.2$, 
showing a portion of a
  geodesic.\label{cmetps}}
\end{center}
\end{figure}

Note that, in accordance with a remark in \cite{Chong:2004hw} that
the Hamilton-Jacobi equation and the massless wave equation admit
separation of variables  for the metric (\ref{cmetric}).

We shall now show that  the existence of a photon surface 
persists in the presence
of cosmological constant and electric field, and for other
static generalizations of the C-metric \cite{Dowker:1993bt,
Emparan:1996ty, Charmousis:2009cm}. These examples show that the appearance of such
surfaces is not restricted to spacetimes of co-homogeneity one, even in the presence of matter.

\subsection{C-metric with Cosmological Constant}
The standard four dimensional ``C-metric'' with cosmological constant and electric charge
may be cast in the form
\ben
ds ^2  = \frac{1}{A^2 (x+y)^2 }\left ( -F(y) dt^2 + \frac{1}{F(y)}dy^2 +
  \frac{1}{G(x)} dx^2 +G(x) d\phi^2 \right) \label{ccmetric}
\een
where
\begin{equation}
F(y) = y^2 + 2mAy^3+e^2A^2y^4-1-\frac{\Lambda}{3A^2}, \qquad G(x) = 1
- x^2-2mAx^3-e^2A^2x^4.
\end{equation}
This solves the Einstein-Maxwell system with field strength $\mathcal{F}=e dy
\wedge dt$. The function $F$ is positive on an interval $(y_0, y_1)$
and the metric is static in this region, with Killing horizons at
$y_0$, $y_1$ corresponding to a black hole horizon and an acceleration
horizon. For sufficiently small $e$, $\Lambda$ the geometry of the static region is essentially the same as for the uncharged C-metric, although the maximal extension is considerably altered \cite{Chen:2015vma}.

The optical metric is given by
\ben
ds_{\rm opt}^2  = \frac{1}{F(y)^2}dy^2 + \frac{1}{F(y)} \left( \frac{dx^2}{G(x)} + G(x) d\phi^2 \right).
\een
After the transformation $y \to -1/r$, this is in precisely
 the form of
equation (4.1) of \cite{Gibbons:2008ru} 
so we see immediately that the projective
structure of the optical metric is invariant under changes of the
cosmological constant. Since $F$ vanishes at the black hole horizon
and the acceleration horizon, it must have at least one maximum on the
interval $(y_0, y_1)$. For small values of $e, \Lambda$, this maximum will be unique. This corresponds to a photon surface, i.e.\ a
totally geodesic submanifold of the optical metric. This surface will
generically have a conical singularity corresponding to that of the
full C-metric.

It is striking that the projective symmetry of the optical metric
first noticed by Islam for the Schwarzschild-de-Sitter metric 
\cite{Islam:1983rxp}
and recently seen to hold
for a wide family of static spherically symmetric solutions
of Einstein's equations \cite{Cvetic:2016bxi} can persist under 
deformations away from spherical symmetry.
Note also that the metric  (\ref{ccmetric}),
is conformal to the metric product of two 2-manifolds each admiting
an isometry.  Thus it shares the property with the standard C-metric that the Hamilton-Jacobi
equation for null geodesics  separates. Since the Ricci scalar is constant, it also follows that
the conformally invariant wave equation separates.     

\subsection{C-metric with conformally coupled scalar field}
In \cite{Charmousis:2009cm} Charmousis et al. construct a
generalisation of the C-metric to allow a magnetic charge and coupling
to a conformally coupled scalar field. The metric takes the form
(\ref{cmetric}) with the metric functions changed to
\begin{equation}
F(y) = y^2 + 2mAy^3+m^2A^2y^4-1-\frac{\Lambda}{3A^2}, \qquad G(x) = 1
- x^2-2mAx^3-m^2A^2x^4.
\end{equation}
The new scalar and electromagnetic field are given by
\begin{equation}
\sqrt{-\frac{\Lambda}{6 \alpha}} \frac{A m (x-y)}{1+A m (x+y)}, \qquad
\mathcal{F} = e dy
\wedge dt+g dx \wedge d\phi.
\end{equation}
Here $\alpha$ is a coupling constant appearing in the action and $g$
is the magnetic charge, related to $e$ and $m$ by
\begin{equation}
e^2+g^2=m^2\left(1+\frac{2 \pi \Lambda}{9\alpha} \right)
\end{equation}

Clearly the modification of $F$ and $G$  does not change the 
conformal and product structure seen in (\ref{cmetric}) and (\ref{ccmetric}).
Thus  we have at least one  photon surface and in addition  the Hamilton-Jacobi
equation for null geodesics  separates. Indeed, for sufficiently small $m, \Lambda$, the polynomial $F(y)$ has four distinct roots\footnote{Note that a degree $p$ polynomial with $p$ distinct roots must have at least one turning point between any two consecutive roots by the intermediate value theorem. Since there are $p-1$ pairs of consecutive roots, and a degree $p$ polynomial has at most $p-1$ turning points, we conclude there is \emph{exactly}  one turning point between any two consecutive roots.}, so in any static region there is at most one photon surface.

\subsubsection{Dilaton C-metric}
The dilaton C-metric of Dowker et al. \cite{Dowker:1993bt} reads:
\begin{eqnarray}
&&ds^2={1\over A^2(x-y)^2}\left[F(x)\left(G(y)dt^2-\frac{dy^2}{G(y)}\right)
+F(y)\left(\frac{dx^2}{G(x)}+G(x)d\varphi^2\right)\right]\nonumber\\
&&e^{-2a\phi}={F(y)\over F(x)},\qquad
A_\varphi=qx, \qquad F(\xi)=(1+r_-A\xi)^{2a^2\over (1+a^2)}\nonumber\\
&&G(\xi)=\bar{G}(\xi)(1+r_-A\xi)^{(1-a^2)\over (1+a^2)},\qquad
\bar{G}(\xi)=\left[1-\xi^2(1+r_+A\xi)\right].\\
\label{dcmetric}
\end{eqnarray}
The region between the horizons satisfies $G(y)<0$, $G(x)>0$ so that
the metric is static with respect to $\partial/\partial t$ and has
optical metric
\ben
ds ^2_{\rm opt} = \frac{dy^2}{G(y)^2} - \frac{F(y)}{G(y)} 
\left( \frac{dx^2}{G(x)
    F(x)} + \frac{G(x)}{F(x)} d\varphi^2\right) \label{dilcopt}
\een
Between the black-hole and the acceleration horizons, $F(y) G(y)^{-1}$ has an extremum so that there is a photon
surface whose geometry is given by the part of the metric in brackets
in (\ref{dilcopt}). Provided $a$ is sufficiently small, this extremum is unique, so there is at most one photon surface in the static patch.

Note that (\ref{dcmetric}) is conformal to the  product metric  
\ben
ds ^2 =  \frac{1}{F(y)}\left(G(y)  dt^2-\frac{dy^2}{G(y)}\right)
+\frac{1}{F(x)} \left(\frac{dx^2}{G(x)}+G(x)d\varphi^2\right)\,.\label{prod}
\een
It again follows that the Hamilton-Jacobi
equation for null geodesics  separates.

\subsubsection{$U(1)^n$ charged C-metric}
Another generalisation of the C-metric, due to Emparan 
\cite{Emparan:1996ty} involves coupling extra $U(1)$
fields and scalars. The appropriate Lagrangian is
\begin{equation}
\mathcal{L} =  R-{1\over 2 n^2}
\sum_{i=1}^n\sum_{j=i+1}^n(\partial\sigma_i-\partial\sigma_j)^2
-{1\over n}\sum_{i=1}^n
e^{-\sigma_i} F_{(i)}^2,
\end{equation}
where the scalars satisfy 
\begin{equation}
\label{singular}
\sum_{i=1}^n\sigma_i =0.
\end{equation}

The C-metric solution is then given by 
\begin{eqnarray}
\label{nonextc}
ds^2 &=& {1\over A^2(x-y)^2}\biggl[ F(x)
\left( {G(y)\over  F(y)}
dt^2-
{F(y)\over G(y)}dy^2\right)+
 F(y)\left({F(x)\over G(x)}dx^2 +
{G(x)\over F(x)}d\varphi^2\right)\biggr],\nonumber
\\
{}\nonumber
\\
A_{(i)\;\varphi}&=& q_i x{\sqrt{(1+r_0/q_i)(1-q_i^2 A^2)}
\over f_i(x)^{n/2}}. \nonumber
\end{eqnarray}
where
\begin{eqnarray}
\label{cmetricn}
F(\xi) &=& \prod_{i=1}^n f_i(\xi),
\qquad f_i(\xi)= (1-q_i A\xi)^{2/n},
\\
e^{-\sigma_i} &=&
{f_i(x)^nF(y)\over f_i(y)^n F(x)}, \qquad G(\xi) = (1-\xi^2)(1+r_0A \xi).
\nonumber
\end{eqnarray}
We take $q_i>0$ and $r_0A>1$. In the region $-1/r_0A<y<-1$, the metric
is static with respect to $\partial/\partial t$. The Killing horizons
at $y=-1/r_0A$ and $y=-1$ are the black hole horizon and the
acceleration horizon respectively. The optical metric takes the form
\begin{equation}
ds ^2_{\rm opt} = \frac{F(y)^2}{G(y)^2}dy^2 -
\frac{F(y)^2}{G(y)} \left(\frac{dx^2}{G(x)} +
\frac{G(x)}{F(x)}d\varphi^2 \right). \label{uncmetopt}
\end{equation}
Note that $G(y)<0$ in this coordinate range. This has again at least
one photon surface for a constant value of $y$ located in the interval
$(-1/r_0A,-1)$ where $F(y)^2/G(y)$ has an extremum. It appears that
this photon surface is unique  for sufficiently small $q_i$. The geometry of
the photon surface is that of the metric in brackets in
(\ref{uncmetopt}). 

Note that Emparan's metric is conformal to one of the form 
(\ref{prod}) and hence the Hamilton-Jacobi
equation for null geodesics  separates.
\subsection{The Melvin Universe and  anti-photon cylinders}

The Melvin universe \cite{Melvin:1965zza} is an electro-vac spacetime which is supported by a homogeneous magnetic field. A uniqueness property is established in \cite{Hiscock:1980zf}, see also \cite{Gibbons:1990um}. It has the spacetime metric
\ben
ds ^2=   G^2(\rho) \Bigl\{ - dt ^2 +  
d \rho  ^2     + d z ^2 \Bigr \}
+ {\rho ^2 \over G^2(\rho) }  d \phi ^2 \,, \een
with 
\ben
G(\rho)= 1 + {B^2 \over 4} \rho^2    
\een
and satisfies the Maxwell-Einstein equations with electromagnetic field
\ben
F = \frac{B \rho}{G^2(\rho)} d\rho \wedge d\phi,
\een
corresponding to a homogeneous magnetic field aligned along the $z$-axis. The optical metric has line element
\ben
ds_{opt.} = d \rho ^2 + d z^2 + \frac{\rho ^2}{G^4(\rho)} d \phi ^2.
\een


The function $\rho^{2} G(\rho)^{-4}$ has a maximum at $\rho = \rho_0:=2/( |B|\sqrt{3})$. Thus the cylindrical surface $\rho = \rho_0$ has vanishing second fundamental form and is therefore totally geodesic. In other words, geodesics initially satisfying $\rho = \rho_0$, $\dot{\rho}=0$ remain tangent to $\rho = \rho_0$. Moreover, any null geodesic in the surface $\rho = \rho_0$ with $\dot{\phi} \neq 0$ is stable, in the sense that a small perturbation will remain close to $\rho = \rho_0$. Null geodesics in the surface with $\dot{\phi} = 0$ are marginally stable, since there are null geodesics with $\dot{\phi} = 0$, $\dot{\rho} = c\neq 0$. Thus $\rho = \rho_0$ is an anti-photon surface, with the conventions of \cite{Cvetic:2016bxi}. Interestingly, this is in contrast to the spherically symmetric case of Reissner--Nordstr\o m, metric with mass $M>0$ and charge $Q$. In the sub-extreme case, $|Q|< M$, there is a unique photon sphere outside the horizon and for the super-extreme case, where $M< |Q| < \frac{3}{2 \sqrt{2}} M$, there is both a photon and an anti-photon sphere \cite{Claudel:2000yi}. 

In \cite{Astorino:2012zm} a generalisation of the Melvin universe to include a cosmological constant is constructed. The metric is modified to:
\ben\label{Astorino}
ds ^2=   G^2(\rho) \left\{ - dt ^2 +  
d z  ^2     + \frac{d \rho ^2}{H(\rho)} \right \}
+ {H(\rho) \over G^2(\rho) }  \frac{\rho^2 d \phi ^2}{1-\frac{\Lambda}{B^2} } \,, \een
with $G$ as previously defined and
\ben
H(\rho) = 1 - \frac{\Lambda}{3} \left( \frac{3}{B^2} + \frac{3 \rho^2}{2} + \frac{B^2 \rho^4}{4} + \frac{B^4 \rho^6}{64} \right).
\een
The electromagnetic field strength becomes:
\ben
F = \frac{B^2}{\sqrt{B^2 - \Lambda}} \frac{ \rho}{G^2(\rho)} d\rho \wedge d\phi.
\een
The $\Lambda \to 0$ case reduces to the Melvin universe above. The $B \to 0$ limit is singular, however after a coordinate transformation the metric can be shown to be equivalent in the $\Lambda <0$ case to a vacuum anti-de Sitter solution found by Bonnor \cite{Bonnor:2008zz}. Both metrics are (up to a coordinate transformation) equivalent to a Horowitz-Myers AdS Soliton \cite{Horowitz:1998ha}. 

One can verify that, provided $-3B^2 < \Lambda < B^2$, the spacetime (\ref{Astorino}) contains an anti-photon surface located at:
\ben
\rho=\rho_0 := \frac{2}{B} \sqrt{\frac{B^2 - \Lambda}{3 B^2 + \Lambda}}.
\een
in the $\Lambda \to 0$ limit, we recover the anti-photon cylinder of the Melvin universe.

Finally, there are also exist anti-photon cylinders in the dilaton-Melvin \cite{Gibbons:1987ps},  \cite{Dowker:1993bt}  metrics. Using (3.2) of \cite{Dowker:1993bt}, the optical metric is:
\ben
ds^2_{\mathrm{opt.}}= dz^2 + d \rho ^2 + \frac{\rho ^2 d \phi ^2 }
{ \bigl( 1+ \frac{(1+a^2) B^2}{4} \rho ^2  \big) ^{\frac{4}{1+a^2} } }\,. 
\een
If the dilaton-photon coupling constant $a$ satisfies $a^2 <3$ there is a unique value of $\rho$ at which 
\be
\frac{\rho^2}{ \bigl( 1+ \frac{(1+a^2) B^2}{4} \rho ^2  \big) ^{\frac{4}{1+a^2} } }
\ee
has a maximum, and hence the situation is the same as for the Melvin universe.

\section{Comments} 

The examples given above may be compared with various
uses  in the literature of the term ``photon sphere''.
Firstly the word ``sphere'' seems inappropriate
since it could be construed to mean a 2-surface which has the 
intrinsic geometry of a round or canonical sphere.
A less misleading  term is ``photon surface''. 
In the case of a  static metric, the most natural definition
would be a totally geodesic submanifold of the optical manifold.
As such, it need not be a level set of the  lapse function
$N$. Indeed in the case of the vacuum C-metric 
\ben
N = \frac{a f(y)}{x+y}\,, 
\een  
which depends upon \emph{both} $x$ and $y$, while the photon surface
is at a fixed  value of $y$. For the Melvin universe, the lapse is constant on the anti-photon surface.

The  definition given  above is much less  restrictive than that used in
several recent uniqueness results 
\cite{Cederbaum:2014gva,Cederbaum:2015aha,Cederbaum:2015fra,Yazadjiev:2015hda,Yazadjiev:2015jza,
Yazadjiev:2015mta,Rogatko:2016mho}
where it is insisted that a photon sphere be a level
set of $g_{tt}$ and any electrostatic potentials. 
A recent attempt has been made to remove that restriction \cite{Yoshino:2016kgi}
and  we suggest therefore, at least in the static situation, that   the term photon surface be limited to that used in the present paper.

Another distinction to be borne in mind 
is  that from  what Teo \cite{Teo} calls  
``Spherical photon orbits around a Kerr black hole''. 
He finds a family of orbits which  lie in a surface of constant $r$ 
in a certain coordinate  system but the surface  is not geometrically a sphere
and moreover not every  photon orbit whose initial tangent lies in the sphere
remains in the sphere.

\section*{Acknowledgements}

The authors are grateful for helpful correspondence from Marco Astorino, Greg Galloway, Carla Cederbaum and Stoytcho Yazadjiev. GWG would like to thank Miriam Cvetic, Chris Pope and Marcus Werner for useful conversations. The work of GWG was partially supported by a LE STUDIUM Professorship held at the University of Tours.

\end{document}